\documentclass[3p,times,twocolumn]{elsarticle}

\usepackage{ecrc}

\volume{00}

\firstpage{1}

\journalname{Nuclear Physics B Proceedings Supplement}

\runauth{K. Kampf}

\jid{nuphbp}

\jnltitlelogo{Nuclear Physics B Proceedings Supplement}

\usepackage{amssymb,amsmath,epsfig}

\usepackage[figuresright]{rotating}

\begin{document}

\begin{frontmatter}



\dochead{\small 5th Joint International Hadron Structure '11
Conference, Tatransk\'a \v{S}trba\vspace*{-0.6cm}
\begin{flushright}
LU TP 11-36
\end{flushright}}

\title{Odd sector of QCD}


\author{Karol Kampf}

\address{Department of Astronomy and Theoretical Physics, Lund
University,\\S\"olvegatan 14A, SE 223-62 Lund, Sweden.}
\address{Charles University, Faculty of Mathematics and Physics,\\
V Hole\v{s}ovi\v{c}k\'ach 2, Prague, Czech Republic\vspace*{-0.8cm}}

\begin{abstract}
A systematic study of the odd-intrinsic parity sector of QCD is presented. We
briefly describe different applications including $\pi^0\to\gamma\gamma$ decay,
muonic $g-2$ factor and test of new holographic conjectures.
\end{abstract}

\begin{keyword}
Chiral Lagrangians\sep 1/N Expansion\sep QCD


\end{keyword}

\end{frontmatter}


\section{Introduction}

The low-energy domain of quantum chromodynamics (QCD) is well understood and
described by an effective field theory called chiral perturbation theory (ChPT). The
derivative ($\rightarrow$momenta$\rightarrow$energy) expansion together with
the symmetry pattern of the underlying QCD organize the inner structure of the
effective Lagrangian and describe the dynamics of the associated Goldstone
bosons (GB). The most important ingredient, the so-called chiral symmetry, the
approximate global symmetry of QCD which acts independently on left- and
right-handed light flavour quarks gives rise to two variants of ChPT. One can
distinguish two-flavour ChPT (active degrees of freedom are $u$ and $d$ quarks) and
three-flavour ChPT (added $s$ quark as well). Their small but not negligible mass
(specially in the case of $s$ quark) is treated as a perturbation. For further
details we refer to the original papers \cite{Gasser:1983ygGasser:1984gg} and
\cite{Bijnens:1999sh} in the case of the next-to-leading order (NLO) and
next-to-next-to-leading order (NNLO), respectively. The number of GB (and
in the two-flavour ChPT $G$-parity) is strictly conserved in
this Lagrangian and we
are talking about even-intrinsic parity sector. At this moment, however, the QCD symmetry pattern cannot
be complete as it would not be possible to describe e.g. $KK\to 3\pi$ or $\pi\to2\gamma$, i.e. well established and existing processes.  The missing odd-intrinsic parity sector arises as a consequence
of the famous chiral anomaly, i.e. of the well-known fact that the dynamics can be invariant under the chiral
symmetry even if operators are not, provided that their variation is a total
derivative \cite{Wess:1971yu}. The explicit construction at NLO can be found in
\cite{Bijnens:2001bb}. More general discussion on chiral anomaly was presented
in the talk of O.~Teryaev at this conference (cf. also \cite{Klopot:2010ke}).

The number of monomials in the above constructed Lagrangian either in even or
odd intrinsic sector increases with higher orders. Information on low-energy
constants (LEC) that stands in front of these monomials thus reflects
the predictivity of ChPT. One possibility how to gain an insight into
them is to use the phenomenology above the domain of ChPT applicability ($\sim
1$
GeV). The first attempt of systematic description of the active degrees of
freedom of the low-lying meson resonances with spin $\leq 1$ was provided in
\cite{Ecker:1988te} for NLO and in \cite{Cirigliano:2006hb} for NNLO, in both
cases for even intrinsic parity sector. This effort was concluded in
\cite{Kampf:2011ty} also for the anomalous sector.

In this work we will focus only on the odd-intrinsic parity sector. We will make a
brief overview of the meson-resonance construction of \cite{Kampf:2011ty} together with so-called resonance saturation and focus on few applications, namely $\pi^0\to\gamma\gamma$ decay and $VVP$ Green function.

\section{Odd-basis construction}

In order to obtain the basis one should first construct all possible hermitian
operators using the standard chiral building blocks and resonance fields
keeping in mind $C$, $P$ and chiral invariance. In order to minimize the set
one uses the partial integration, equation of motions, Bianchi identities,
Schouten identity and basic relations among the chiral building blocks.
The constructed Lagrangian of this meson-resonance odd-parity sector becomes:
\begin{equation}
\mathcal{L}_{R\chi T}^{(\text{ odd})}= \varepsilon^{\mu\nu\alpha\beta}
\sum_{X,i}\kappa _{i}^{X}%
\mathcal{O}_{i\,\mu\nu\alpha\beta}^{X}\,.  \label{reslag}
\end{equation}%
The $X$ stands for the resonances entering the individual terms, linear in
resonances: V (18 monomials), $A$ (16), $S$ (2), $P$ (5), quadratic: $VV$ (4),
$AA$ (4), $VA$ (6), $VS$ (2), $VP$ (3), $AS$ (2), $AP$ (2), and finally three
trilinear monomials: $VVP$, $VAS$, $AAP$. All 67 monomials can be found in
\cite{Kampf:2011ty}. The contribution of $\kappa_i^X$ introduced in
(\ref{reslag}) to LEC will be discussed in the next section.

\section{Resonance saturation}

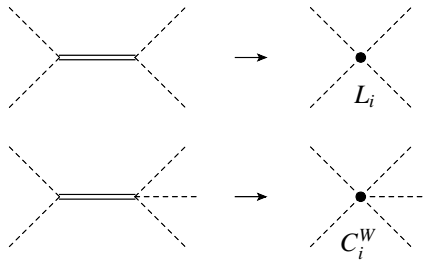
\begin{figure}[h]
\begin{center}
\includegraphics{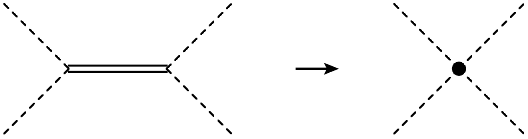}
\begin{picture}(0,0)(0,0)
\put(-25,3){$L_i$}
\end{picture}
\\[0.4cm]
\hspace*{1.4mm}\includegraphics{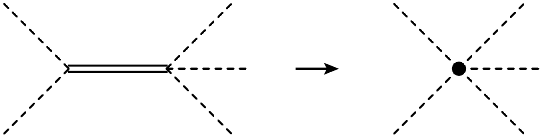}
\begin{picture}(0,0)(0,0)
\put(-34,0){$C_i^W$}
\end{picture}
\end{center}
\caption{Schematic description of resonance (double line) saturation for both even intrinsic sector (NLO LEC denoted by $L_i$) and anomalous sector (LEC represented by $C_i^W$).}
\label{schem}
\end{figure}

The resonance saturation both for even-intrinsic parity and odd-intrinsic parity is schematically depicted in Fig.~\ref{schem}.
LEC are thus expressed by
means of resonance parameters $\kappa_i^X$. One usually uses a term saturation
provided we neglect all other contribution. In principle one has exchanged one
set of parameters with another. In the latter case we can use, however, richer
phenomenological information which includes explicit resonances.
In the construction of resonance monomials in~(\ref{reslag}), the large $N_C$
approximation was tacitly assumed. As so all LEC  can be saturated except that
they are subleading in large $N_C$. 
In the anomalous NLO sector we have altogether 23 LEC constants from which 21
can be saturated schematically as
\begin{equation}\label{Cwiab}
C^W_i = a_i N_C^2 + b_i N_C + O(N_C^0)\,.
\end{equation}
Let us note that the formal large $N_C$ enhancement ($a_i\neq 0$ for $i=6,8,10$)
is due to $\eta'$ exchange. The explicit form for all $a_i$ and $b_i$ can be
found in \cite{Kampf:2011ty}.
A systematic elimination of $\kappa_i$ in~(\ref{Cwiab}) leads to one
relation which includes only LEC and the resonance parameters from
the even sector
\begin{equation}
\frac{F_V^2}{2 G_V} C^W_{12} = F_V (C^W_{14} - C^W_{15}) + G_V C^W_{22}\,.
\end{equation}

\section{Decay $\pi^0 \to \gamma\gamma$}
Studying the odd-intrinsic parity sector of QCD one cannot avoid the discussion
on the most important anomalous process $\pi^0\to\gamma\gamma$. This decay
was crucial in establishing the role of the anomaly for the gauge theory
\cite{Adler:1969gkBell:1969ts}. Without the anomaly it was a long standing
puzzle how to overcome the implication of the Sutherland theorem which stated
that this process should be suppressed by power of $m_\pi^2/(1\,\text{GeV}^2)$
to its actual measured value. A new experiment PrimEx \cite{Larin:2010kq} with a
total uncertainty of 2.8\% (with planned improvement) motivates even to better
understand these Sutherland contributions. Presumably accidental non-existence
of the leading logarithm (i.e. terms $\sim m_\pi^2  \log m_\pi^2$)
\cite{Donoghue:1986wvBijnens:1988kx} leads to a necessity to calculate even
higher orders in $m_\pi$. In the language sketched in Introduction one needs to
calculate this process up to NNLO in ChPT. This was performed for two-flavour
ChPT in \cite{Kampf:2009tkKampf:2009zz} with remarkable simple analytic result.
In order to use a phenomenological information the transition to three-flavour
ChPT must be employed
\cite{Kampf:2009tkKampf:2009zz,Gasser:2007sg,Gasser:1983ygGasser:1984gg}. Apart
from  LO and logarithms one ends up with two LEC $C_7^W$ and $C_8^W$ on
which $\pi^0 \to \gamma\gamma$ depends. The first one can be set using
the resonance
phenomenology as explained in the previous section. The second one can be
obtained from the decay $\eta\to\gamma\gamma$: here we have similar dependence
on $C_7^W$ and $C_8^W$ as for $\pi^0$ decay. However, this process is known
only at the NLO and again it contains no leading logarithms. First attempt to
calculate higher-order logarithms can be found in \cite{Bijnens:2010pa}.
Assuming these corrections would be small we may
obtain~\cite{Kampf:2009tkKampf:2009zz}
\begin{equation}
 \Gamma_{\pi^0\to\gamma\gamma} = (8.1\pm 0.1)\,\text{eV}\,.
\end{equation}

\section{Renormalization group equation and leading logs}
In the previous section we were discussing the most significant process of the
odd sector $\pi\to \gamma\gamma$ and mentioned an important concept of the
leading logarithms. These logarithms were obtained calculating two-loop
diagrams without any simplifications. It is however well known that the leading
logs (up to all orders) can be obtained from one-loop diagrams only
\cite{Weinberg:1978kzBuchler:2003vw}. This enables to open program of
calculating the leading logs to a sufficiently high order estimating general
prescription and eventually resum the series. It should in any case help to
understand the connection of Taylor expansion in ChPT and asymptotic series of
QCD \cite{Bakulev} in some intermediate region. This program was
started with massive $O(N)$ non-linear sigma model in
\cite{Bijnens:2009ziBijnens:2010xg}. Among other things the authors calculated
the physical mass and decay constant $F_\pi$ up to five-loop order. As
$O(N+1)/O(N)$ for $N=3$ is isomorphic to $SU(2)\times SU(2)/SU(2)$ i.e. to two-flavour ChPT
(without isospin breaking) we can use the same program to make further
prediction also in the anomalous sector for two flavours. This program is in
progress \cite{BKL}.

\section{$VVP$ Green function}

The previous example $\pi^0\to\gamma\gamma$ can be obtained from a three-point
Green
function involving two vector currents and one pseudoscalar density, in short
$VVP$. As this function can be further related to other interesting physical
quantities we will first discuss some of its general properties. Let us start with the
definition
\begin{equation}\label{defVVP}
\Pi_{\mu\nu}^{abc}(p,q) =\!\! \int \!\! d^4x\, d^4y\, e^{ip\cdot x+iq\cdot y} \langle
0| T[V^a_\mu(x)V^b_\nu(y)P^c(0)|0\rangle,
\end{equation}
with
\begin{equation*}
V_\mu^a(x) =\bar q(x) \gamma_\mu \frac{\lambda^a}{2}q(x)\,,\qquad P^a(x)
=\bar q(x) \mathrm{i}\gamma_5 \frac{\lambda^a}{2}q(x)\,.  \label{VPcur}
\end{equation*}
This object can be calculated using the Lagrangian~(\ref{reslag}). Nine of
odd-intrinsic parity resonance parameters $\kappa_i$ contribute:
$\kappa^V_{12,14,16,17}$, $\kappa^{VV}_{2,3}$, $\kappa^P_5$, $\kappa^{PV}_3$
and $\kappa^{VVP}$. We can assume that the result should correctly describe
also high-energy region which is accessible directly by perturbative QCD
(procedure known as operator product expansion (OPE)). Using this link we arrive
to the result which depends only on two unknown parameters $\kappa^{VVP}$ and $\kappa_3^{PV}$.
Our theoretical construction can be connected with an existing phenomenological
model known as LMD+P \cite{Moussallam:1994xp} (for LMD see below).
Here we will write the result only in the limiting on-shell case (working in
the chiral limit it means
$r^2=0$) when one parameter drops out and we have a simple dependence only on
one
of them:
\begin{multline}\label{Pircht}
\frac{r^2}{B_{0}}\Pi ^{\text{R}\chi \text{T}}(p^{2},q^{2};r^{2}\to 0)=
\frac{1}{(p^2-M_V^2)(q^2-M_V^2)} \\
\times \Bigl\{\tfrac12(p^2+q^2)(32\sqrt2 d_m F_V \kappa_3^{PV} + F^2) - \frac{M_V^4 N_C}{8\pi^2}\Bigr\}\,.
\end{multline}
Before performing OPE we can assume the construction of $VVP$ using the vector resonances only (i.e. dropping pseudoscalar resonances) and connecting the currents only via lowest
lying vector resonance. This procedure is known as vector meson dominance (VMD). It can be achieved in~(\ref{Pircht}) by setting $\kappa_3^{PV}$ to
\begin{equation}
\kappa_3^{PV} \,\stackrel{\text{VMD}}{=}\, - \frac{F^2}{32\sqrt2 d_m F_V}\,,
\end{equation}
so the first term on the second line of~(\ref{Pircht}) cancels out. Such a
fine-tuning is, however, not natural in our formalism and we prefer to fix its
value from phenomenology.
Defining the on-shell pion-$\gamma^*\gamma^*$ formfactor
\begin{equation}
{\cal F}_{\pi \gamma^*\gamma^*}(p^2,q^2) = \frac{2}{3 B F}\lim_{r^2\to0}r^2 \Pi(p^2,q^2;r^2)
\end{equation}
we can connect the measurements of this formfactor with the value of
$\kappa_3^{PV}$ in~(\ref{Pircht}). Using the combined fit on CLEO and BABAR
data \cite{Gronberg:1997fjAubert:2009mc} we get
\begin{equation}\label{k3pv}
 \kappa_3^{PV} = -0.047\pm 0.018\,.
\end{equation}
From information on $\pi(1300)$ decay modes we can set \cite{Kampf:2011ty} also
the value for the last parameter entering $VVP$
\begin{equation}\label{kVVP}
 \kappa^{VVP} = (-0.57 \pm 0.13)\,\text{GeV}\,.
\end{equation}

\subsection{g-2}
Having study $VVP$ correlator it is impossible to avoid discussion on its role
in the famous muon $g-2$ factor. One of the major systematic error in
determining this factor lies in the so-called hadronic light-by-light (LbL)
scattering. This has several contributions, one of which can be related to $VVP$
correlator defined in~(\ref{defVVP}). Using the values in~(\ref{k3pv})
and~(\ref{kVVP}) we can obtain a prediction for the $\pi^0$-exchange in
hadronic LbL scattering:
\begin{equation}
a_\mu^{\text{LbyL;}\pi^0} = (65.8 \pm 1.2)\times 10^{-11}\,,
\end{equation}
which is the most precise up-to-date determination of this quantity, very close to the
recent AdS/QCD studies in \cite{Cappiello:2010uy}.

\section{AdS/QCD and anomaly}

A direct connection between low energy QCD and its high energy version in the
odd-intrinsic parity sector using new holographic methods
\cite{Maldacena:1997re} was suggested recently in
\cite{Son:2010vc}\footnote{Note that after this work was presented a new study
on the anomalous $VVA$ in the soft-wall holographic model has appeared
\cite{Colangelo:2011xk}.}.
Within a wide range of holographic models Son and Yamamoto have established a
relation which connects the longitudinal and transverse part of the $VVA$ (it
can be directly connected with $VVP$ of the previous section) and two-point
correlator $\langle LR\rangle$
\begin{equation}\label{SYrel}
w_L(Q^2) - 2 w_T(Q^2) = -\frac{2 N_C}{F^2} \Pi_{LR}(Q^2)\,,
\end{equation}
valid for all $Q^2$.
This relation was tested in pQCD domain with not very satisfactory result
\cite{Knecht:2011wh}. We will address here its validity for a small $Q^2$
within a non-perturbative regime of QCD. Using the language of LEC we can
rewrite~(\ref{SYrel}) into
\begin{equation}\label{c22l10}
C_{22}^W = - \frac{N_C}{32 \pi^2 F^2} L_{10}\,,
\end{equation}
which connects the odd-intrinsic parity LEC ($C_i^W$) with even parity LEC
($L_i$) at NLO. Using the resonance saturation discussed in Section 3 (for
$L_{10}$ see \cite{Ecker:1988te}) we start with
\begin{align}
C_{22}^W &= -\frac{F_V \kappa _ {17}^V}{\sqrt{2} M_V^2}-\frac{F_V^2 \kappa _
3^{\text{VV}}}{2 M_V^4}\,,\\
L_{10} &= \frac{F_A^2}{4 M_A^2} - \frac{F_V^2}{4 M_V^2}\,.
\end{align}
Employing OPE of the previous section together with further asymptotic
behaviour (namely Weinberg sum rule and OPE on the axial pion form factor) we
end up with a condition
\begin{equation}
 k^{PV}_3 \,\stackrel{\text{AdS/QCD}}{=}\, \frac{M_V^2 N_C - 16 \pi^2
F^2}{1024\pi^2 d_m F} \approx 0.019\pm 0.01\,,
\end{equation}
which is numerically close to zero\footnote{As an interesting outcome (but not
directly connected to the studied problem) the last equation suggests that
$M_V = \frac{4\pi F}{\sqrt{N_C}}$ which might be a consequence
of an asymptotic behaviour for some so-far unstudied correlator.} and thus in a
disagreement with the phenomenological study (cf.(\ref{k3pv})). On the other
hand
$k_3^{PV}=0$ is just a result of the so called LMD (lowest meson dominance)
model \cite{Peris:2000twKnecht:2001xc} (i.e. where one drops systematically $P$). It is
thus interesting to notice that the studied class of holographic models produces
a relation~(\ref{c22l10}) which seems to be valid in the LMD model.

\section{Conclusion}

A systematic study of the odd-intrinsic parity sector was presented. It
concludes similar study in the even sector. Variety of applications was briefly
described using different methods. We have shown a general concept of the
resonance saturation for LEC in the odd sector; showed its implication for the
important decay of
this sector:$\pi^0 \to \gamma\gamma$ decay and sketched further program for it
using renormalization group equations.
The general $VVP$ Green function and its phenomenological applications were
shortly discussed. The last section
was devoted to the modern holographic study focused on this sector with a
particular check.

\section*{Acknowledgment}
I would like to acknowledge the work of my collaborators: Johan Bijnens,
Stefan Lanz, Bachir Moussallam, Ji\v{r}\'{\i} Novotn\'y and Jaroslav Trnka, as
well as the organizers of this conference.



\nocite{*}
\bibliographystyle{elsarticle-num}



\end{document}